# Embedded System to Detect, Track and Classify Plankton Using a Lensless Video Microscope

Thomas G. Zimmerman, Vito P. Pastore, Sujoy K. Biswas, Simone Bianco

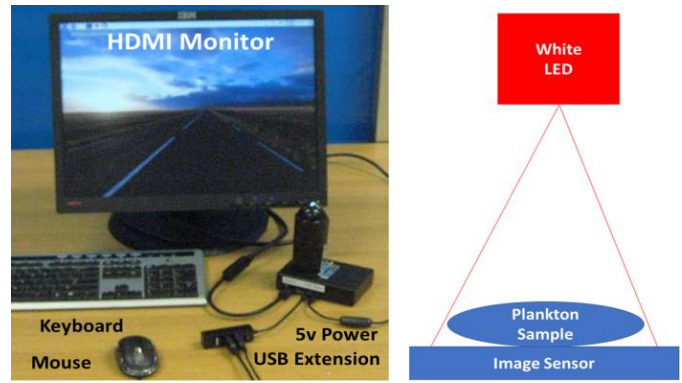

Fig. 1. (Left) Lensless microscope system build with a Raspberry Pi 3. (Right) A white LED 100 mm above the image sensor casts shadows of plankton swimming on top of the image sensor's protective glass cover.

*Abstract*— Plankton provide a foundation for life on Earth. To advance our understanding of the marine ecosystem for scientific, commercial and survival purposes, more *in situ* continuous monitoring and analysis of plankton is required. Cost, complexity, power and data communication demands are barriers to widespread deployment of *in situ* plankton microscopes. We address these barriers by building and testing a lensless microscope with a data pipeline optimized for the Raspberry Pi 3. The pipeline records 1080p video of multiple plankton swimming in a sample well while simultaneously detecting, tracking and selecting salient cropped images for classification @ 5.1 frames per second. Thirteen machine learning classifiers and combinations of nine sets of features are evaluated on nine plankton classes, optimized for speed (F1=0.74 @ 1 msec. per image prediction) and accuracy (F1=0.81 @ .80 sec.). System performance results confirm that performing the entire data pipeline from image capture to classification is possible on a low-cost open-source embedded computer.

## I. Introduction

Plankton are vital for life on Earth, providing most of the breathable oxygen, carbon sequestration and larvae nutrition for the planet. Plankton are any organism that live in water and are unable to swim against current. This broad definition includes bacteria, phytoplankton and zooplankton and organisms like starfish that are only plankton during their larval (juvenile) stage of life. In this paper we are primarily interested in plankton within our imaging range (~50-1000 um).

To understand the environmental and biological processes that regulate plankton populations, a fundamental need is to count and classify species over time and space [1]. This is a daunting endeavor considering 71% of the Earth's surface is water-covered. Satellite cameras can view phytoplankton communities by detecting chlorophyll concentrations using blue and infrared wavelengths [2]. But to advance our understanding of the marine ecosystem more *in situ* continuous monitoring and analysis is required [3][4].

### A. Microscopes

The development of the microscope in the 1600's helped established the field of microbiology. Organisms smaller than the eye could finally be seen, their morphology captured in drawings and their identity organized by observable traits. Observing the movement and interactions of plankton can advance our understanding of essential activities such as reproduction, eating and predator avoidance.

Traditionally biological samples are brought to the microscope. The Continuous Plankton Recorder (CPR) Survey is the longest continuous sampling of plankton in open waters. Since 1931 commercial ships traveling in the North Atlantic and North Sea have volunteered to tow a one-meter long metal box, at a depth of 5 to 10 meters, to collect and preserve plankton samples. The box containing bands of filter silk that mechanically advances as the box is pulled through the water, trapping and preserving plankton samples[2]. The boxes are sent to a laboratory in Plymouth, England where the silk is cut into strips and distributed to staff who count large zooplankton by eye and phytoplankton by microscope [5].

The simple and robust mechanical design and opportunistic trips on commercial vessels has contributed to the CPR's long-term success. Since its inception, CPRs have surveyed over 6.6 million nautical miles. Over 258k samples have been analyzed

---

This work is funded by the National Science Foundation (NSF) grant No. DBI-1548297. Disclaimer: Any opinions, findings and conclusions or recommendations expressed in this material are those of the authors and do not necessarily reflect the views of the National Science Foundation; TGZ (tzim@us.ibm.com), VPP (vitopaolopastore@gmail.com), SKB (skbhere@gmail.com), SB (sbianco@us.ibm.com)

[2] http://www.antarctica.gov.au/about-antarctica/environment/climate-change/biology/continuous-plankton-recorder

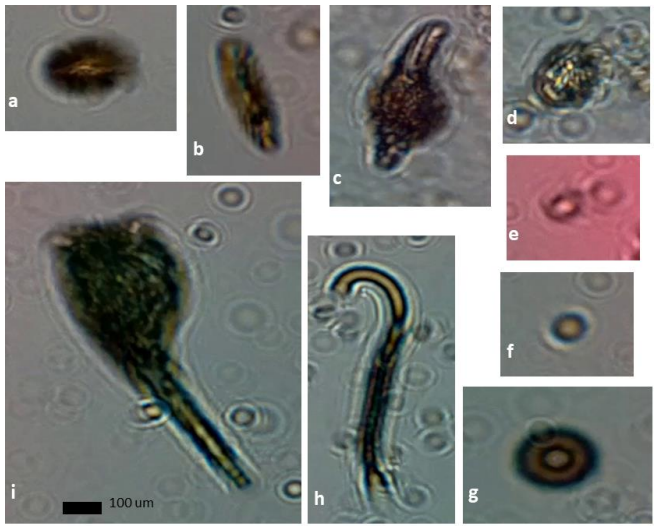

Fig. 2. Sample training images for 9 classes. All images are displayed at the same scale. (a) Didinum (b) Paramecium (c) Blepharisma (d) Euplotes (e) Algae1 (f) Algae2 (g) Volvox (h) Dileptus (i) Stentor

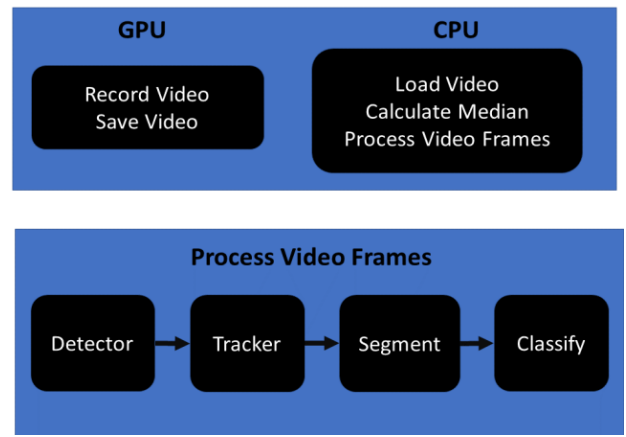

Fig. 3. Data processing pipeline. (Top) The GPU records and saves videos while simultaneously the CPU loads the last saved video, calculates a median image from 20 randomly selected frames, then processes the video frames. (Bottom) Each module processes the entire video and passes results to the next module. The Detector reduces image resolution to speed up detection and outputs a list of object bounding boxes. The Tracker assigns object ID's by inter-frame proximity and outputs a list of tracklets. For each tracket, the Segment module selects the object with the largest aspect ratio (indicating a profile view), extract features and performs a classification.

and classified into 698 taxa. The samples provide a near-century longitudinal study of plankton population dynamics linking marine environmental changes to global warming [6]. However, the labor-intensive sample processing limits the volume of data that is recovered [7].

With the advent of electronic cameras and video recorders, underwater microscopes can now the shape and motion of live plankton in their natural habitat [8]. Underwater microscopes can be classified by the method used to deliver samples into the imaging region. The three most common methods are open field, passive and active flow [9]. The open field method utilizes a still or video camera fitted with an extreme macro lens with a long working distance. The camera is brought close to the subject and reflected light is collected and magnified by the lens [10].

In a passive flow microscope, the camera faces a light source separated by an open gap (shadowgraph imaging), allowing plankton to drift into the imaging volume [11]-[14]. In both open field and passive flow methods, plankton are free to swim, though their motion is dominated by ambient water currents.

Active flow microscopes draw or pump samples into a vessel with an optical path between a light source and a camera. In one of the most complex examples, a flow cytometer injects water containing plankton through a narrow optical viewing chamber, surrounded by a sheath fluid (typically water, often including a surfactant). By carefully adjusting the flow rates of the two fluids, hydrodynamic focusing is achieved, separating plankton samples, like "pearls on a string", aligning them down the center of the viewing chamber.

The Imaging FlowCytobot[3] (IFCB) is a submersible flow cytometer for automatic imaging of plankton [15]. The device exemplifies many positive traits of an *in situ* plankton microscope. The IFCB produces up to 8.3 high quality greyscale images per second (1380x1034, 3.4 pixels/micron), is commercially available and provides an end-to-end solution.

Open-source software is available to collect, distribute, classify, display and share data[4]. However, several traits impede its widespread deployment.

The IFCB is a complex exquisitely designed electrical, mechanical and optical scientific instrument. The device consumes 35 watts, weighs ~32 kg, costs ~$150,000[5], requires replacement fluids (biocide, bleach and sheath fluids), has limited component lifetime (motors and syringes with a 1 to 2 year lifetime), captures a static monochromatic image, requires complex setup and maintenance and does not perform on-board feature extraction or classification[6]. The captured cropped images can be locally stored on a 512 GB solid-state drive and uploaded by cable over Ethernet.

*B.  Plankton Detection*

When an object enters the IFCB microscope it triggers a xenon flashlamp and captures an image of a single plankton on a monochromatic CCD camera. The short flashlamp duration (1 usec) minimizes blurring to ~ 1 um as the object flows through the viewing chamber. Hydrodynamic forces exerted by the laminar sheath fluid flow align the object's major axis with the camera's longer axis, providing a consistent pose, simplifying classification [15].

The Video Plankton recorder (VPR) is a towed passive flow microscope that continually samples objects that pass through its optical aperture [16]. A 10 us strobe synchronized to the video camera's frame rate minimizes blurring. Unlike a flow cell, objects can have any orientation and position. Image quantization is performed with a fixed threshold. Edge strength is used to detect in-focus objects that are cropped and saved. Small objects below a threshold area are rejected. The quantization, edge strength and minimum object area thresholds

---

[3] https://mclanelabs.com/imaging-flowcytobot
[4] https://github.com/hsosik/ifcb-analysis/wiki
[5] https://ucscsciencenotes.com/feature/detecting-deadly-algae
[6] https://mclanelabs.com/imaging-flowcytobot/ifcb-manuals-0/

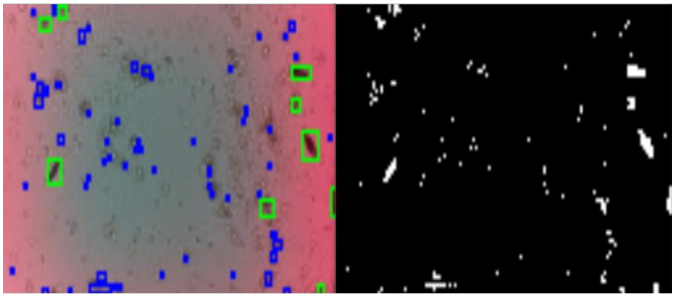

Fig. 4. (Top) Frame from captured video at full resolution (1080p). Green box indicates object with acceptable area. Blue box indicates small rejected object. (Bottom) Frame resized by factor of 15 (128x72) to reduce detection time and remove tiny objects.

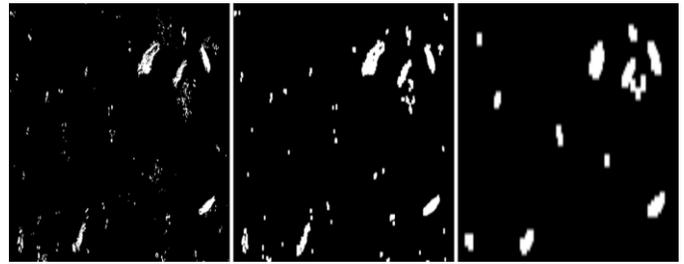

Fig. 5. Demonstration of how reducing frame resolution removes tiny objects. Full scale color image is converted to grayscale, binary quantized with a fixed threshold, then resized. From left to right; full captured resolution (1920x1080), 1/5 resizing (384x216) and 1/15 resizing (128x72). The 1/15 resizing is used in the Detection module.

are imperially set.

Bi, et al. [17] analyzes plankton with the ZOOVIS underwater microscope, a towed passive flow microscope with a large (30 cm) depth of field and ~240 mL sampling volume per frame. A red LED pulsed at 5 us is synchronized to the 15 Hz frame rate of a 5 Mpix CCD sensor. The illumination is not uniform, making it difficult to use a global threshold for image quantization. Instead, they combined two segmentation strategies; a Maximally Stable Extremal Regions technique to capture large planktons (~0.5 mm2) and an adaptive threshold based on Otsu's method to capture small plankton.

### C. Tracking

Tracking plankton in *situ* creates datasets to enable biologists to study plankton motion, to gain a deeper understanding of marine ecosystems [18]-[24]. Zooplankton migrate to the surface at night when it is safe from predators to feed on phytoplankton, then descend into deeper water during the day, where they excrete their carbon-rich diet. This "carbon pump" captures CO2 from the air and deposits it on the seabed, sequestering carbon, mitigating global warming. Predator-prey interaction is vital to maintain population balance. Disrupting the swimming behavior of one species can have dramatic effects on the environment, giving rise to harmful algal blooms

The natural currents and induced water flow in the three *in situ* microscope sample delivery methods do not provide favorable conditions for studying plankton movement. Instead, plankton swimming dynamics are studied in the lab in stationary vessels of water with multiple cameras [25].

Tracking multiple objects in video is a well-studied topic in computer vision [26]. Tracking is simplified when there are few objects in the image, minimizing object overlap [27]. A tracklet is a sequence of locations of a single object across multiple frames, produced by continually tracking a detected object. However, a detector may occasionally miss an object (detector dropout), giving the appearance that the object disappeared, resulting in an object being treated as a new object when it reappears.

When two objects cross paths, merging and splitting events occur causing identification confusion, a data association problem [28]. Merging and splitting can be detected when proximal objects in a previous frame show dramatic changes in their properties [29]. Several strategies are used to preserve ID's during these events [30]. For example, if two merged objects have significantly different identifiable traits (e.g. area) once splitting occurs, their ID's can be re-established based on their distinguishing traits.

### D. Classification

The quantities of images produced by digital microscopes combined with the development of image processing and machine learning tools has led to a growing number of efforts to perform supervised classification of plankton [31]-[33].

The 2015 Kaggle "Plankton Challenge"[7] attracted over 1,000 teams, producing over 15,000 entries that performed supervised classification on 121 classes from a dataset of 50 million silhouette images. The images were collected by the Hatfield Marine Science Center using a towed ISISS passive flow microscope [14]. The winning entry used a convolutional neural network (CNN) with over 27 million parameters trained over 24 hours on several NVIDIA GPUs[8]. The feature set included image size and shape moments (Hu and Zernike), Linear Binary Patterns and Haralick texture features.

Classification by CNN has been benchmarked on the Rasberry Pi 3 using the MobileNet v2 SSD model and Tensor Flow Light libraries[9]. Optimized for embedded devices, the model classifies 300 x 300 pixel images from the Common Objects in Context (COCO) data set with an average inference time of 272 msec.

Zheng *et al*. [9] improves on previous classifiers by optimizing features and using a combination of kernel functions. Ten feature sets are applied to three plankton image datasets (WHOI, ZooScan and Kaggle) to determine the optimal combination to produce the highest classification accuracy. The datasets have 22, 20 and 30 classes, respectively. A performance baseline is established with an SVM classifier using a Gaussian kernel, producing F1 scores of 0.88, 0.82 and 0.77, respectively. By combining linear, polynomial and Gaussian kernel functions, classification F1 accuracy increases to 0.90, 0.88 and 0.84, respectively.

### E. Plankton Monitoring Ecosystem

The IFCB is an end-to-end plankton monitoring system. By making the image processing and classification code open-source, scientists and developers can share, use and improve plankton image data and analysis. The commercial availably of the IFCB provides a means to replicate and deploy a standard high-quality device for collecting *in situ* plankton images.

---

[7] https://www.kaggle.com/c/datasciencebowl/data
[8] https://github.com/benanne/kaggle-ndsb/blob/master/doc.pdf
[9] https://www.hackster.io/news/benchmarking-tensorflow-and-tensorflow-lite-on-the-raspberry-pi-43f51b796796

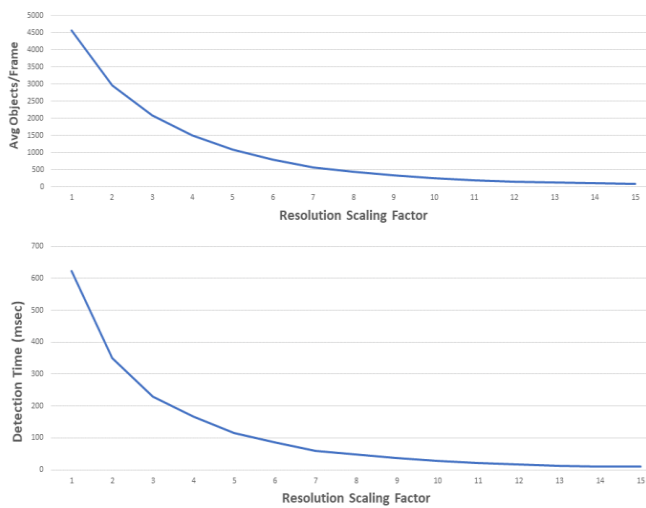

Fig. 6. (Top) Average number of objects detected in a frame for the Euplotes training video as a function of resizing factor. (Bottom) Average processing time for all objects in a frame as a function of resizing factor.

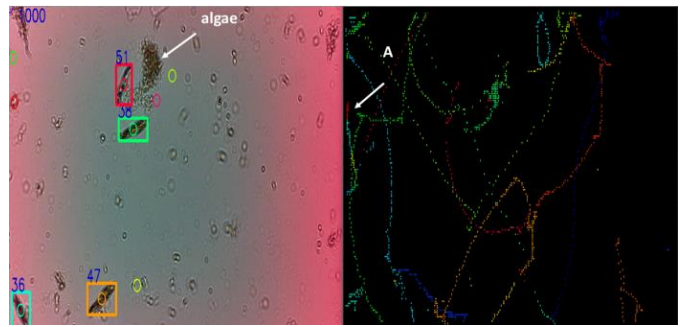

Fig 7. (Left) Eight objects are detected (circle) in this video frame. Four of them are currently being actively tracked (rectangular box with ID's). The clump of algae to the right of ID 51 is rejected by the Detector module. (Right) Location trails of nine objects integrated over 1000 frames. Each trail color represents a tracklet of an object with a unique identification (ID). Each dot is a position, determined at 30 times a second (video capture frame rate). The trail ending at label "A" is created by an object that goes in and out of the frame, creating short tracklets (red, cyan, yellow, blue, green) that are rejected by the Segment module for being too short.

The image capture and region-of-interest cropping occurs in the IFCB device. The rest of the processing occurs remotely. To manage the tremendous amount of data of image data (terabytes) an IFCB can generate during a long deployment (months), a dashboard interface[10] is provided to organize, summarize, search and present images and metadata.

Due to the high cost, IFCBs are usually deployed temporarily for specific missions and cruises, often to study harmful algae blooms[11]. Reducing the cost of *in situ* microscopes several orders of magnitude (e.g. >$1000) would substantially increase deployment opportunities[12]. This would introduce a new set of challenges; how to transmit, distribute and process billions of plankton images produced by thousands of *in situ* microscopes, distributed across the globe, often located in remote areas?

We recognize this as an IoT problem where devices on the edge (microscopes) produce a massive amount of data (plankton images), placing an extreme burden on the network and end users. The solution is to use an edge computing approach; embed the processing (classification) in the sensor (microscope). The microscope performs data compression by sending descriptors (species labels, features and statistics) instead of raw data (image pixels). This represents a data collection paradigm shift from image datasets from a few individual site-specific expeditions to aggregated reports from a wide distributed network of continuously sampling *in situ* microscopes for ecosystem model building, assessment and forecasting [34][35].

In this paper we present the following methods developed to perform image capture, processing and classification on a Raspberry Pi 3 embedded system;

1) **Image Capture and Pre-Processing**. Use the Pi's video co-processor (GPU) and ARM processors (CPU) to simultaneously capture and process video, respectively. Reduce image resolution to decrease detection time by removing objects too small to classify.
2) **Detection.** Use a median filter to create uniform lighting to enable the use of a global quantization threshold. Apply dilation to avoid one object being detected as several. If several objects appear in the cropped image, select the one with the largest area.
3) **Tracking.** Use an object persistence model to compensate for detector dropout. Detect merge and split events as large changes in inter-frame object area.
4) **Feature Extraction.** Optimize a feature set based on a classification accuracy vs. computation tradeoff. For each tracklet, perform feature extraction on one object, selected with the largest aspect ratio, indicating a profile view.
5) **Classification.** Detect and remove algae images before feature extraction and classification. Create classes for algae that are not removed. Select the machine learning (ML) algorithm with the optimal accuracy vs. computation tradeoff.

## II. METHODS

### A. Lensless Microscope

A hybrid microscope was developed that combined open field and active flow, where plankton are drawn into a viewing chamber without a sheath fluid, left to settle so there are no currents, then a video is recorded so all motion is due to plankton activity. This procedure is simulated in the lab by hand pipetting plankton samples into the microscope sensing well, allowing the water to settle, then recording a video. Heat from the image sensor did create some noticeable convective flow in the water after prolonged use, causing some otherwise non-motive plankton, like algae, to move.

---
[10] https://github.com/joefutrelle/ifcb-dashboard/wiki/User-guide
[11] https://mclanelabs.com/category/ifcb/
[12] https://ucscsciencenotes.com/feature/detecting-deadly-algae/

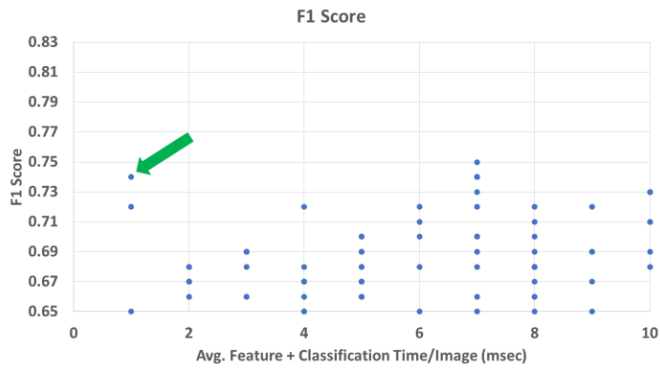

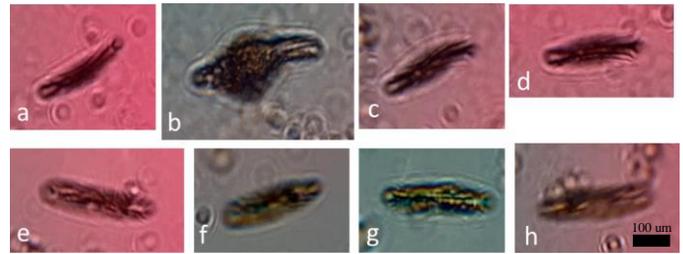

Fig. 8. Each point is the F1 score of a combination of 9 feature sets and 13 classifiers. The x axis is the average time to calculate the features and classify one image. The plot is used to determine the optimal tradeoff of computation time and classifier accuracy. The green arrow points to the Decision Tree Classifier using geometric and area features, considered to be the optimal tradeoff of computation time and accuracy. Classification requires an average of 1 msec. per image. The feature computation time is considered zero since the features are calculated and used by the Detector and Tracker modules.

Fig. 9. Four training images of Blepherisma (top row) and Paramecium (bottom row). Images are to scale. The midline bulge in the anatomy of the Blepherisma (b), due to ample ingestion of food, is captured in some training images but is not consistently present in training and test images. When absent, Blepherisma is often misclassified as Paramecium.

The system uses a lensless microscope [36] built with commodity hardware to reduce cost and complexity (Fig. 1). The data pipeline performed object detection and tracking of multiple objects (plankton) captured at HD 1080p resolution (1920 x 1080 @ 30 fps). Classification algorithms and feature sets were selected to optimize the tradeoff of computation speed and accuracy (F1 score). The data pipeline ran on an unmodified Raspberry Pi 3 without any additional hardware accelerator support.

The lens was removed from a Raspberry Pi Camera V1. A PVC card (0.8 mm thick) with a 12 mm square punched hole was adhered with aquarium-safe silicone sealer over the image sensor's protective glass cover, creating a plankton sample well. A single white LED was mounted at the top of a 100 mm black 38 mm diameter PVC tube, approximating a point light source, casting shadows of plankton onto the image sensor.

The image sensor (OV5647, 3.76 x 2.74 mm, 1.4 um pixel, 2592x1944 pixels) was connected to the Pi's high-speed Camera Serial Interface (CSI). No electronic modifications were made, so the image sensor operated as a Pi camera, which allowed the use of Pi camera libraries and utilities.

### B. Plankton Samples

Seven live plankton species (Fig. 2) were purchased[13] to create a data set of images. For each plankton species at least two short videos (71 or 144 seconds) were recorded; one for training and one for testing. A 50 uL solution of plankton was hand pipetted onto the image sensor well and allowed to settle for 10 seconds. The recordings were made using the Pi OS rapsivid utility with the default settings (automatic gain, 1920 x 1080 resolution, 30 frames per second, H.264 encoding, ~ 1.6 MB/sec file size). The videos were converted to MP4 format with the open-source FFmpeg utility to enable random access to frames.

---

[13] Protozoa Set Item #131020, Carolina Biological Supply Company, Burlington, NC

A library of 24 videos was created. Each video featured one species, although contamination with algae, tiny ciliates and detritus was common. The videos were reviewed by a human observer for quality. Some videos were rejected for excessive concentration of plankton, a rare condition *in situ*, or for lacking the target species. Of the remaining videos, seven videos were selected for training and seven for testing. None of the videos were edited.

### C. Training and Test Set

A collection of salient images was automatically generated from the seven training videos using the data pipeline described in this paper. Salient images are defined as having a large aspect ratio, indicating a profile view. Thirty-three images were manually selected as supervised training images for each of the seven plankton classes.

There were a considerable number of salient images of algae, indicating a failure of the algae failure in the detector due to motion (caused by thermal currents from image sensor heat) and/or smooth texture (few edges). The algae images that passed through the detector had morphology was distinct enough to sort them into two group. We defined two new classes (*Ag1* and *Ag2*) by selecting 33 images from each of these two algae group. The final supervised selection process produced a total balanced training set of 9 classes with 33 members per class (Table II).

The training set was relatively small due to the limited number of short videos we captured, too small for neural network (NN) classifiers. In our initial selection of classification algorithms to run on the Raspberry Pi 3 platform, we excluded NNs due to their long inference time, compared to other non-NN supervised ML algorithms (e.g. instance-based, decision tree and Bayesian). In addition, NN training time is extremely long on a Raspberry Pi, a factor of concern for our long-term goal of developing a network of AI-microscopes that can perform distributed learning, where each embedded microscope performs unsupervised learning on the plankton they locally encounter.

The seven test videos were passed through the data pipeline (Fig. 3), to create a collection of 1237 salient test images. Each image was assigned one of ten categories by a human observer; one for each of the nine classes (Table I1) and one for a rejection group. The rejection group consisted of 54 images (4.4%) of objects that the human observer could not identify, typically unspecified algae or algae that obscured a species beyond recognition. The final unbalanced test image set consisted of 1183 labeled images. The labels were required to measure the accuracy of the classifier predictions and were not revealed to the classifiers during testing.

TABLE I
CONFUSION MATRIX FOR DTC BEST TIME (L) AND BEST F1 (R)

|     | ble | did | dil | eup | par | ag1 | ste | ag2 | vol |
| --- | --- | --- | --- | --- | --- | --- | --- | --- | --- |
| ble | 41  | 0   | 3   | 9   | 44  | 0   | 0   | 0   | 0   |
| did | 0   | 82  | 0   | 1   | 0   | 0   | 0   | 0   | 16  |
| dil | 7   | 0   | 89  | 2   | 0   | 0   | 0   | 0   | 0   |
| eup | 3   | 5   | 0   | 74  | 0   | 11  | 0   | 2   | 1   |
| par | 46  | 0   | 3   | 5   | 37  | 5   | 0   | 0   | 0   |
| ag1 | 0   | 0   | 0   | 24  | 0   | 73  | 0   | 1   | 0   |
| ste | 0   | 0   | 0   | 0   | 0   | 0   | 100 | 0   | 0   |
| ag2 | 0   | 0   | 0   | 0   | 0   | 32  | 0   | 67  | 0   |
| vol | 0   | 3   | 0   | 2   | 0   | 0   | 0   | 0   | 93  |

|     | ble | did | dil | eup | par | ag1 | ste | ag2 | vol |
| --- | --- | --- | --- | --- | --- | --- | --- | --- | --- |
| ble | 61  | 1   | 0   | 1   | 36  | 0   | 0   | 0   | 0   |
| did | 0   | 96  | 0   | 0   | 0   | 0   | 0   | 3   | 0   |
| dil | 3   | 0   | 95  | 0   | 0   | 0   | 0   | 0   | 0   |
| eup | 0   | 19  | 0   | 57  | 4   | 16  | 0   | 2   | 0   |
| par | 0   | 5   | 0   | 0   | 86  | 5   | 0   | 1   | 0   |
| ag1 | 0   | 7   | 0   | 1   | 0   | 80  | 0   | 10  | 0   |
| ste | 0   | 0   | 0   | 0   | 0   | 0   | 100 | 0   | 0   |
| ag2 | 0   | 0   | 0   | 0   | 0   | 21  | 0   | 78  | 0   |
| vol | 1   | 1   | 0   | 0   | 1   | 0   | 0   | 0   | 96  |

TABLE II
TRAINING AND TEST DATA SET

| Class | Species | Train Images | Test Images |
| --- | --- | --- | --- |
| blep | *Blepharisma* | 33 | 328 |
| eup | *Euplotes* | 33 | 169 |
| did | *Didinium* | 33 | 177 |
| par | *Paramecium* | 33 | 158 |
| dil | *Dileptus* | 33 | 132 |
| vol | *Volvox* | 33 | 82 |
| ag1 | *Algae1* | 33 | 58 |
| ste | *Stentor* | 33 | 51 |
| ag2 | *Algae2* | 33 | 28 |
| Total Images | | 297 | 1183 |

D. Image Processing

1) Resolution Reduction

Video was recorded at full resolution (1920x1080@ 30 fps) in order to provide high resolution salient images for feature calculations (Fig. 4a). However, the object bounding box coordinates used for image cropping did not require high resolution (Fig. 4b). Reducing image resolution reduced detection computation time and image fragmentation, a condition where one object was detected as multiple objects.

Very small objects (e.g.<50 um) appeared as halos due to light diffraction and do not present enough spatial features for classification. Reducing frame resolution before image quantization fortuitously caused small objects to disappear (Fig. 5), reducing the number of objects in a frame (Fig. 6a) and object detection time (Fig. 6b). This is particularly useful in plankton microscopy where small plankton tend to be more numerous in nature [37].

We empirically determined a resolution reduction factor of 15 retained enough small objects of interest while substantially reducing the tiny unclassifiable objects. For example, the range of average total objects per frame detected at full resolution for the training videos was from 503 (*Stentor*) to 7687 (*Euplotes*). The latter video produced an exceptionally large number of detected objects due to the excessive concentration of tiny algae, detritus and fragmentation of large objects. When the frame resolution was reduced by a factor of 15, the average total objects detected per frame was reduced to 45 and 354, respectively.

Objects were segmented in the reduced resolution frame by intensity quantization using a fixed threshold, determined empirically. The same fixed threshold was used for all the training and testing videos. If the object's intensity dynamic range straddles the threshold, object fragmentation could occur, resulting in one object being detected as multiple objects. This is often addressed by applying dilation and erosion operations to the binary quantized image to fill the small holes in the disjointed object. We observed that reducing image resolution using area averaging (cv2 INTER_AREA method) with an integer scaling factor, produces a similar desired effect of reducing object fragmentation.

2) Correcting for Uneven Lighting

The microscope's point light source creates uneven lighting across the image sensor, forming a bright center that decreases radially. An illumination correction was performed for each frame by subtracting a reference image, to enable the use a global threshold, a fast and simple method to binary quantize a high contrast image. The reference image was created once per video by calculating the median intensity from 20 randomly selected frames (hence the need for MP4 video format). Each selected frame was scaled by a factor of 15 and converted to grayscale. We found that selecting the red channel of the color array was faster and provided better contrast than using the RGB calculation (cv2.COLOR_BGR2GRAY). The first 100 frames of the video were ignored to give the image sensor's automatic gain control (AGC) time to settle.

E. Detector

The detector sequentially processed each low resolution (128x72) frame, detecting and saving the coordinates of objects within an acceptable area range of the 9 classes (1k to 40k pixels).

Each frame was loaded, converted grayscale by selecting the red channel. The resolution was reduced by a factor of 15 and a fixed threshold was applied to the illumination-corrected frame to produce a binary image. The area and coordinates of each object's bounding rectangle were determined with the OpenCV findContours method.

Objects within the acceptable area were dilated with a kernel size of seven, then the findContours method was applied again. If more than one object was detected, the object with the largest area was selected. The contour of the final selected object was filled with the drawContours method, creating a solid blob object without holes suitable for feature extraction.

During developed a source of significant classification error was contamination by algae. Algae are a diverse group of plankton that take on many shapes, as cells join in large amorphous groups. To address this challenge, we developed an algae detector to remove algae images from the pipeline before feature extraction and classification. Algae were detected as objects with a high texture (determined by summing a Canny edge detected image) and little movement, a benefit of tracking

TABLE III
FEATURE SETS AVG. CALCULATION TIME ON A RASPBERRY PI

| Key | Classifier |
|---|---|
| KNN | Nearest Neighbor |
| NCT | Nearest Centroid |
| DTC | Decision Tree |
| RFC | Random Forest |
| GNB | Gaussian Naïve Bayes |
| LDA | Linear Discriminant Analysis |
| QDA | Quadratic Discriminant Analysis |
| LSV | Linear Support Vector Machine |
| NSV | Nu-Support Vector |
| BNB | Bernoulli Naïve Bayes |
| ABC | Ada Boost |
| GBC | Gradient Boosting |
| ETC | Extra Trees |

TABLE IV
CLASSIFIERS TESTED ON PLANKTON DATA SET

| Key | Feature Set | Features | Calculation Time (msec) |
|---|---|---|---|
| g | Detector geometry | 4 | 0 |
| a | Area | 1 | 0 |
| G | Geometric | 14 | 2.2 |
| h | Hu moments | 7 | 0.3 |
| z | Zernike moments | 25 | 218.6 |
| L | local binary patterns | 54 | 572.8 |
| H | Haralick intensity | 13 | 193.3 |
| i | Intensity histogram | 8 | 3.2 |
| f | Fourier descriptors | 10 | 72.4 |

plankton in substantially stationary water.

*F. Tracking*

The tracking module received a list of objects from the detector and created tracklets; a sequence of positions for each detected object (Fig. 7). Each tracklet was assigned a unique label (ID) that persisted across frame boundaries. A tracklet begins when an object is first detected and terminates when it leaves the viewing area or encounters another object (merging). A tracket also begins when a merged object separates (splitting).

*1) Inter-Frame Object Matching Using Proximity*

For each frame a list was created of Euclidian distances between each object in the current frame and all objects in the previous frame. The list was sorted in ascending order and each current object was matched to the closest unassigned object in the previous frame. Once an object was matched, it was considered assigned. If the distance was too great (an empirical threshold), the object in the current frame was assigned a new ID. Once this matching process was complete, any unassigned objects were considered "ghost" objects and retained their ID.

*2) Temporal Persistence for Long-Term Object Tracking*

During detection, an object may temporarily disappear if its brightness drops below the quantization threshold. When this occurs, the ghost object is still there, just not detected. When it reappears, it would be treated as a new object, creating multiple ID assignments [38], an undesirable outcome.

To address this condition, each ghost object was assigned an age counter that would increment for each frame it was unassigned. If the age counter exceeded a threshold (10 frames, empirically determined), the object would be deleted (removed from further tracking consideration). If the object reappeared within the age threshold, it would rejoin the other active tracked objects. The age threshold must be carefully tuned based on the density of objects and the frequency of detector dropout. Setting the threshold too low results in "blinking", assigning new ID's to the same object. Setting the threshold too high creates "land mines", assigning the ghost's ID to an object that passes over the ghost's location.

*3) Merge and Split Detection*

Merging occurs when two objects come in proximity and are detected as one object. Splitting occurs when merged objects move apart and are detected as two objects. Merging and splitting cause the loss and creation of an ID, respectively. Since the frequency of these events increases with object density and plankton density in nature is often relatively low, we elected to implement a simple strategy.

Merging and splitting was detected when the relative area of an object's bounding box across a frame boundary dramatically increases and decreases, respectively. The threshold was established by observing the statistics of inter-frame bounding box area changes for the training video with the highest density of objects. A tracklet was rejected when merging occurs. A new tracket with a new ID's was created when splitting occurs.

*G. Segment*

The segment module selects one salient image per tracklet. A salient image is defined as the image of an object in a tracket with the largest aspect ratio that does not touch the image boarder. We observed that when objects touch the image boarder, they often contained two overlapping objects or a big mass of algae, so we added the condition. Image border contact was detected by summing the binary image boundary. If the sum was greater than zero, the image was rejected.

Long trackets increase the probability that a rotating object will present a pose favorable for classification. Trackets with less than 10 frames (empirically determined) were rejected to reduce fragment traces caused by successive merging and splitting. Tracklets that do not show substantial movement were rejected, as they are often algae or detritus. For the seven training videos, on average 68% of the trackets were sufficiently long, while only 24% were sufficiently long and moving.

Each accepted salient image was cropped from the high-resolution color frame using the object's low-resolution rectangular bounding box coordinates provided by the tracker.

*H. Feature Extraction*

Nine sets of features were evaluated for their contribution to the accuracy (F1 score), training and testing times, for all 13 classifiers (Table III). The first feature set were features calculated for and used by the detect and track modules, so their use in classification did not contribute any additional computational load. They consisted of area (number of on pixels in binary quantized image), aspect ratio (from a tight bounding box), texture (sum of Canny edge pixels divided by area) and contour smoothness (contour pixels divided by square root of area). Area was also evaluated as an independent feature for it is the most common observable trait for binning plankton [39]. The remaining seven feature sets examined shape (Fourier descriptors, geometry, Hu and Zernike movements) and texture (local binary patterns, Haralick, intensity histogram) [40].

TABLE V
AVERAGE PROCESSING TIME PER FRAME

| MODULE | AVG | MIN | MAX |
|---|---|---|---|
| MEDIAN | 0.9 | 0.5 | 1.5 |
| DETECT | 93.3 | 87.5 | 100.1 |
| TRACK | 2.9 | 0.9 | 7.4 |
| SEGMENT | 98.9 | 88.5 | 126.5 |
| CLASSIFY | .007 | .003 | .038 |
| TOTAL | 196 | 177.5 | 235.5 |
| FPS | 5.1 | 5.6 | 4.2 |

TABLE VI
PROCESSING TIME OF SELECTED PROCEDURES

| MODULE | AVG % | PROCEDURE | AVG % |
|---|---|---|---|
| MEDIAN | <1% | | |
| DETECT | 46% | | |
| | | Load Frame | 53% |
| | | Image Processing | 41% |
| | | Detect Objects | 5% |
| TRACK | 1% | | |
| SEGMENT | 49% | | |
| | | Load Frame | 58% |
| | | Process Segments | 37% |
| | | Save Images | 5% |
| CLASSIFY | <1% | | |

To calculate the average computation time required to calculate each feature set on an image on the Pi 3, each feature set was run on all 295 training images (one image failed to produce a complete feature set).

*I. Classifiers*

Thirteen classifiers were selected (Table IV) to represent a diverse group of multi-class supervised learning methods covering discriminant analysis, support vector machines, nearest neighbor, Gaussian process, naïve Bayes, decision tree and ensemble methods. The classifiers are from the scikit-learn.org supervised learning library and were used "out of the box" without any custom tuning.

### III. RESULTS

The modules in the digital pipeline were developed and debugged on a laptop (Intel i7-6600U @ 2.6 GHz, 16G RAM) in Python, then the code and libraries were executed on a Pi 3 to determine the computation times presented in the tables.

*A. Module Execution Time*

The processing times for each module, averaged over all the frames in the seven training videos are listed in Table V. The Median module was run once for each video at the beginning of the data pipeline (min=1.3 sec, avg=2.2 sec, max=2.8 sec), but is presented as an average over the frames in the video, to be consistent with the other metrics.

Table VI presents the average time per frame of significant procedures in each module as a percentage of the total computation time. Over half of the Detector and Segment module's time was spent loading the frame from memory.

The classify time per frame was insignificant because; classification occurred once per valid tracklet, there were on average 48 valid tracklets per video in the training videos (146 frames per classification), the optimal classifier (DCT) required only 1 msec. per image and the classifier used features calculated and used in the Detect and Track modules.

*B. Classifier Evaluation*

The 13 classifiers were run against all combinations of the 9 feature sets. The feature set combination that produced the highest F1 score for each classifier is presented in Table VII. An F1 score of 0.81 was achieved with the LDA classifier using a combination of Zernike moments, local binary patters, Haralick and intensity features, required an average of 988 msec. for feature calculations and 10 msec. for inference, per image. The classifier labeled DTC_OP was the optimal combination of classifier and feature set that produced the best tradeoff of computation time vs. F1 score, 1 msec. and 0.74, respectively, determined by a scatter plot of F1 vs. computation time (Fig. 8). The optimal feature set consisted of features calculated and used in the Detect and Track modules; area, aspect ratio, normalized edge sum and contour smoothness. A confusion matrix for these conditions is presented in Table I (Left). A confusion for the DTC with the feature set combination that produced the highest accuracy (F1=0.81) is presented in Table I (Right).

### IV. DISCUSSION

Our objective was to develop a low-cost microscope and data pipeline that could classify a few species of plankton at video rates. We have demonstrated the detection, tracking and supervised classification of nine plankton classes on an unmodified Pi 3 running on average at 5.1 frames per second with an F1 accuracy of .74 using an inexpensive lensless microscope under a wide range of sample population densities.

We used two performance metrics to guide our system design; video frame processing time and classification F1 score. We utilized the Pi's GPU and CPU by simultaneously capturing video while processing the previous video.

During the initial design phase, we assumed the dominant computational load would be feature calculation and classification based on our prior work with neural networks. The pipeline was designed to select the least number of redundant images (same object) with the highest classification confidence. The design produced the desired effect, performing classification on average once every 146 frames.

TABLE VII
CLASSIFER PERFORMANCE

| Classifier | Features | Acc | P | R | F1 | featureTime | TrainTime | PredictTime | TotalTrain | TotalPredict |
|---|---|---|---|---|---|---|---|---|---|---|
| LDA | zLHi | 0.83 | 0.82 | 0.84 | 0.81 | 988 | 169 | 10 | 1157 | 998 |
| LSV | gzLi | 0.83 | 0.8 | 0.85 | 0.81 | 795 | 754 | 9 | 1549 | 804 |
| NSV | gGLi | 0.8 | 0.78 | 0.77 | 0.77 | 578 | 95 | 211 | 673 | 789 |
| RFC | gaGLHi | 0.79 | 0.77 | 0.8 | 0.77 | 772 | 106 | 21 | 878 | 793 |
| ETC | ghzHi | 0.78 | 0.75 | 0.77 | 0.76 | 415 | 68 | 19 | 483 | 434 |
| KNN | gahzi | 0.78 | 0.75 | 0.77 | 0.75 | 222 | 5 | 402 | 227 | 624 |
| NCT | gahzi | 0.77 | 0.74 | 0.77 | 0.75 | 222 | 4 | 7 | 226 | 229 |
| DTC_OP | ga | 0.74 | 0.73 | 0.75 | 0.74 | 0 | 6 | 1 | 6 | 1 |
| GNB | gai | 0.76 | 0.74 | 0.76 | 0.74 | 3 | 8 | 12 | 11 | 15 |
| BNB | gaGzi | 0.75 | 0.72 | 0.78 | 0.73 | 224 | 7 | 8 | 231 | 232 |
| GBC | gazi | 0.72 | 0.74 | 0.73 | 0.73 | 222 | 6340 | 66 | 6562 | 288 |
| QDA | g | 0.71 | 0.74 | 0.71 | 0.7 | 0 | 8 | 8 | 8 | 8 |
| ABC | zLHif | 0.45 | 0.45 | 0.53 | 0.46 | 1060 | 1460 | 338 | 2520 | 1398 |

Image resolution reduction was initially performed to speed up display frame rate during debugging. When it was observed that small unclassifiable objects would disappear, it was incorporated into the data pipeline, reducing object detection for dense samples.

Feature and classification optimization evaluation revealed the dominant computational load was caused by retrieving images. By sharing images between the Detect and Segment modules an average of 57 msec. per frame could be saved, increasing the average pipeline computational speed from 5.1 to 7.2 frames per second.

In this paper we collected plankton trajectory data, but only used it to select salient images for classification. Future work will examine the use of trajectory profiles to support environmental and behavior research, and as classification features.

Algae presents a significant classification challenge as they can take on many shapes. Limiting detection to moving objects with low to moderate texture eliminated most algae but can inadvertently eliminate classifying other species. Defining two classes for the algae that passed detection worked well except for the algae (ag1) that contaminated the *Euplotes* samples.

The largest source of classification error was the confusion between Blepharisma and Paramecium. The shape of Blepharisma changes based on nutrition [43]. An underfed Blepharisma appears cylindrical (Fig. 9 a,c,d), similar in silhouette to Paramecium (Fig. 9 e-h), while a well-nourished Blepharisma has an expanded posterior (Fig 9 b).

In order to better distinguish the underfed Blepharisma from the Paramecium, more detailed features in the grayscale image must be extracted. These features are obscured by diffraction in the current microscope. Adding a blue laser to the microscope produces holographic images with much higher feature resolution, along with 3D tracking information, for a nominal cost (~$1) [44][45]. Signification computation is required to reconstruct each holographic image, a challenge we will address in a future publication, but worth the effort for digital in-line holography appears to be the most cost-effective method to improve image resolution.

Finally, there are over 4,000 species of plankton in the world [41], making it impractical to create a comprehensive dataset to train a *in situ* microscope to classify every possible species. Fortunately, the variety of species in a specific area (species richness) is low, typically in the range of 10-50 [42]. This simplifies the classification task. Each microscope must only classify the small collection of location-specific species it will encounter. But how could one train thousands of microscopes for the collection of site-specific species each would encounter *a priori*? We postulate the solution to this conundrum is for each microscope to learn *in situ* [40].